\documentclass[preprintnumbers, showkeys, superscriptaddress]{revtex4}  

\usepackage[]{graphicx}

\begin{document} 
\title{Compressive Phase Contrast Tomography}

\author{F. R. N. C.  Maia}
\affiliation{National Energy Research Scientific Computing Center, Lawrence Berkeley National Laboratory, Berkeley, California, 94720, USA}
\author{A. MacDowell}
\affiliation{Advanced Light Source, Lawrence Berkeley National Laboratory,  Berkeley, CA, USA}
\author{S. Marchesini}
\email[Correspondence and requests for materials should be addressed
to S. M.:]{smarchesini@lbl.gov}

\author{H. A. Padmore}
\author{D. Y. Parkinson}
\affiliation{Advanced Light Source, Lawrence Berkeley National Laboratory,  Berkeley, CA, USA}
\author{J. Pien}  
\affiliation{CUDA consultant, www.jackpien.com}
\author{A. Schirotzek }
\affiliation{Advanced Light Source, Lawrence Berkeley National Laboratory,  Berkeley, CA, USA}
\author{C. Yang}
\affiliation{Computational Research Division, Lawrence Berkeley National Laboratory,  Berkeley, CA, USA}


\preprint{LBNL-3899E}

\begin{abstract}
When x-rays penetrate soft matter, their phase changes more rapidly
than their amplitude.  Interference effects visible with high brightness
sources creates higher contrast, edge enhanced images. When the object
is piecewise smooth (made of big blocks of a few components), such
higher contrast datasets have a sparse solution. We apply basis
pursuit solvers to improve SNR, remove ring artifacts, reduce the
number of views and radiation dose from phase contrast datasets
collected at the Hard X-Ray Micro Tomography Beamline at the Advanced
Light Source. We report a GPU code for the most computationally
intensive task, the gridding and inverse gridding algorithm (non
uniform sampled Fourier transform).

\end{abstract}


  \maketitle 
\section{INTRODUCTION}
\label{sec:intro}  
Phase contrast tomography allows the imaging of light materials that would
otherwise be transparent to x-rays, and obtaining edge enhancment at
higher SNR for the same dose \cite{nugent}\cite{paganin}. 
Further dose
reduction is expected from the application of compressive sensing
reconstruction techniques. This is what we set out to do in this paper.

The refractive index of X-rays passing through light materials is very
close to a real number; when written in the in the form:
$n=1-\delta+i\beta $, $\delta$ is one to 3 orders of magnitudes larger
than $\beta$ (depending on the energy of the X-rays); exploiting the
refractive contrast seems obvious.

While it is not possible to observe phases, experimental methods exist
to observe the Laplacian of the phase changes induced by propagation
through an object. By rotating the object around an axis, one can
collect a series of such projections. Standard tomographic processing
of the data yields the laplacian of the object in the three
dimensional space\cite{paganin}.

The phase contrast mechanism is especially useful for defining the
boundaries between composite objects. If the sample is made of large
blocks of  components, the boundaries will be thin and far from
each other. In other words,  the solution will be sparse.

We tested this concept using experimental phase contrast data from 60
micron glass spheres at the tomography beamline 8.3.2 at the Advanced
Light Source, Lawrence Berkeley National Lab. The setup is similar to
standard tomography procedures \cite{BL832} in that samples are rotated in a
monochromatic X-ray beam and the transmitted X-rays are imaged via a
scintillator, magnifying lens and a digital camera to give an
effective voxel size in the reconstructed three-dimensional image of
1.8 $\mu$m. 

\begin{figure}
   \begin{center}
   \begin{tabular}{c}
   \includegraphics[width=\textwidth]{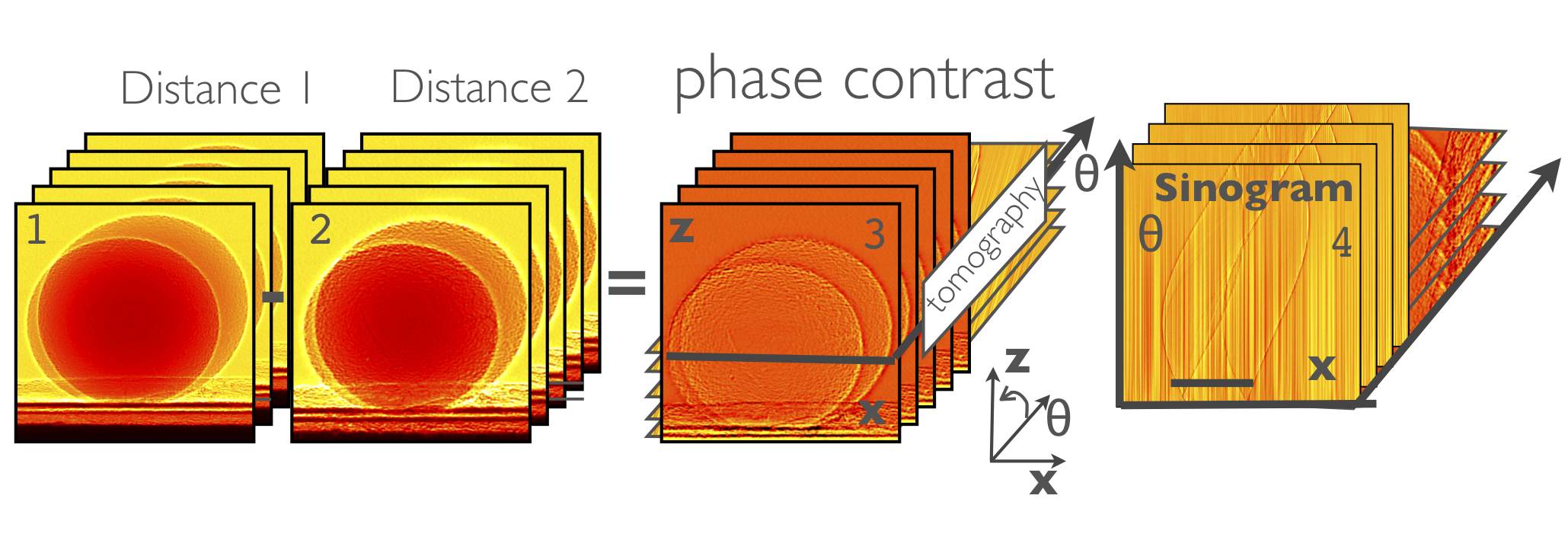}
   \end{tabular}
   \end{center}
   \caption[example] 
   { \label{fig:pctomo} 
Propagation based phase contrast and tomography of two two glass balls placed 
in the microtomography beamline 8.3.2.  From left to right:
(1) projection radiograph at short distance from the sample, (2) same as (1) but with the detector 1 m. downstream. (3) difference between (1) and (2), (4) sinogram.
}
   \end{figure}

Background normalized images are shown in Fig. \ref{fig:pctomo}, along
with their difference showing the propagation based phase contrast
enhancement, and the corresponding sinogram. Significant artifacts arise from
 the residual fluctuations in the illuminating beam due
to vibrations and the high power beam impinging on a crystal
monochromator. Defective detector elements (e.g. dead
pixels in a CCD) with non-linear responses to incoming intensity will
appear in the reconstructions as sharp rings with a width
of one pixel. Similar artifacts also arise from dusty or damaged
scintillator screens. Miscalibrated detector pixels, e.g. due to beam
instabilities not completely taken into account by a normalization
correction, give rise to wider and less marked rings\cite{munch}.

\begin{figure}
   \begin{center}
   \begin{tabular}{c}
   \includegraphics[width=\textwidth]{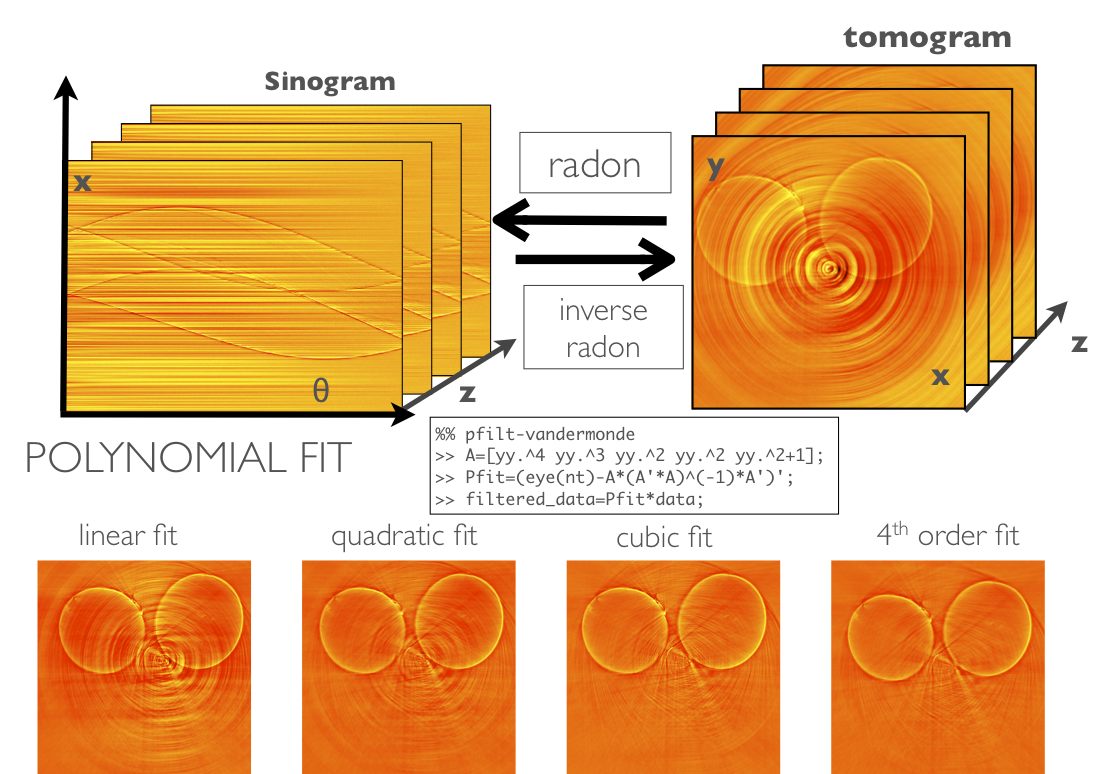}
   \end{tabular}
   \end{center}
   \caption[example] 
   { \label{fig:pfilt} 
 Significant artifact due
to residual fluctuations in the illuminating beam due to vibrations and 
the high power
beam impinging on a crystal monochromator. Comparison of image quality using different filters (polynomial subtraction) (Fig. \ref{fig:pfilt}). }
   \end{figure}

To improve the reconstruction we employed a 3rd
order polynomial fit to smooth the sinogram and reduce the ringing
artifacts.  Increasing the order of the polynomial beyond 3 did not
improve the image further (Fig. \ref{fig:pfilt}).


   \begin{figure}
   \begin{center}
   \begin{tabular}{c}
   \includegraphics[height=6cm]{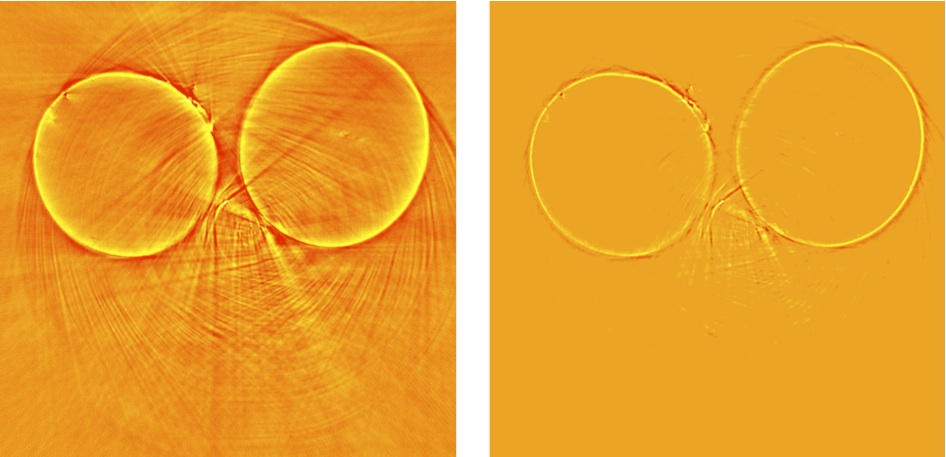}
   \end{tabular}
   \end{center}
   \caption[example] 
   { \label{fig:bp} 
Comparison of the reconstruction using a filter and using SPGL1 basis pursuit solver with the same filter.
}
   \end{figure} 

We formulated the final reconstruction problem as an L1-minimization problem, i.e. 
$$
\min || x ||_1 \,\,\,\,
\textnormal{subject to } ||\textnormal{filter} (\textnormal{Radon} \{ x\} -\textnormal{data})||_2 <\epsilon
$$ where $x$ is the phase contrast image to be reconstructed and
filter is a polynomial interpolation operator designed to reduce the
ringing artifact in the reconstruction, and $\epsilon$ is a
regularization parameter we choose in advance. We solve the
L1-minimization problem by using the SPGL1 software developed by
E. van den Berg and M. P. Friedlander [1]. The software requires us to
provide a function to perform $y =$ Radon $\{x\}$. Due to the large volume
of data produced by at beam line (BL.8.3.2) at the Advanced Light Source,
we would like perform phase contrast tomographic reconstruction in
real time.

   \begin{figure}
   \begin{center}
   \begin{tabular}{c}
   \includegraphics[width=\textwidth]{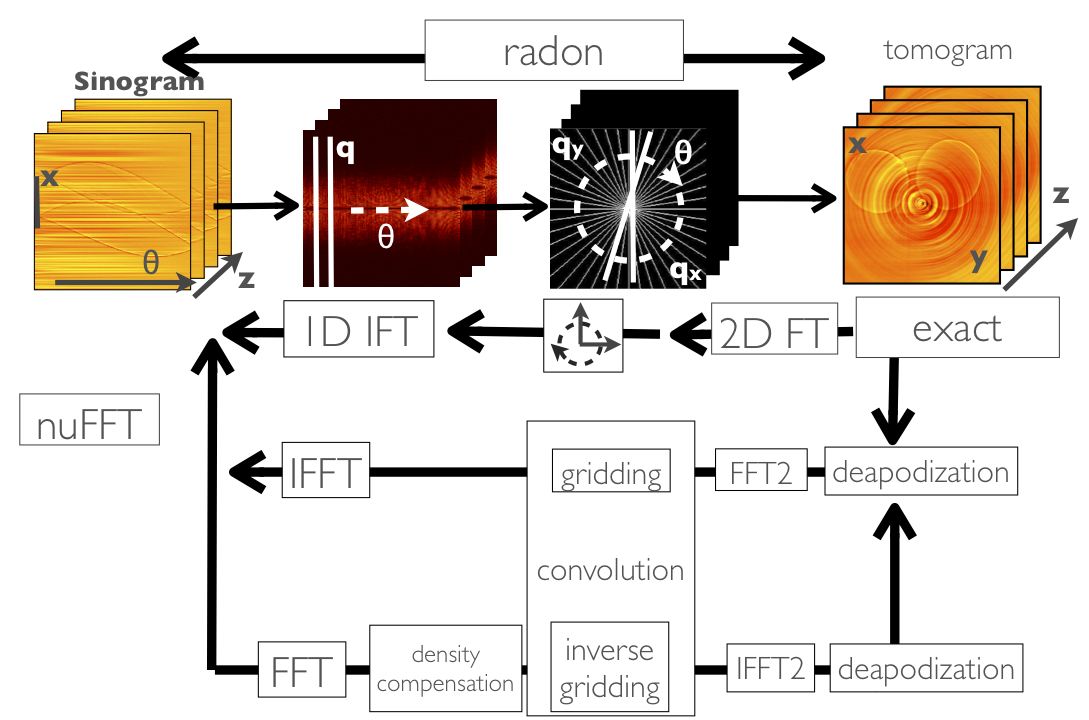}
   \end{tabular}
   \end{center}
   \caption[example \label{fig:gnuradon}] 
   { \label{fig:gnuradon} 
Fast Radon and inverse Radon transforms implemented on a Tesla GPU computing engine.}
\end{figure}

\begin{figure}
   \begin{center} \begin{tabular}{c}
   \includegraphics[width=\textwidth]{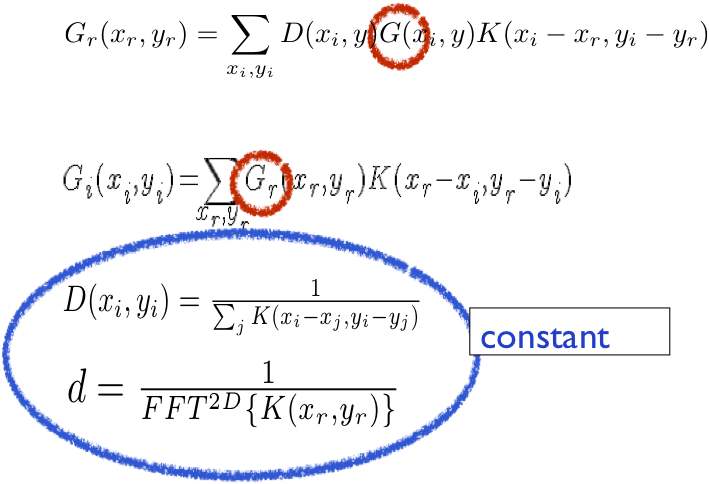} \end{tabular}
   \end{center} \caption[example]
   { \label{fig:gridding} 
Fast Radon and inverse Radon transforms implemented on a Tesla GPU computing engine.}
\end{figure}

\section{fast radon transform}
Fast Radon and inverse Radon transforms form the computational kernels
of many image reconstruction algorithms. The 2D Radon tranform of an
image x can be implemented in a number of ways. One of the most
efficient ways to perform Radon\{$x$\} is to first perform a 2D FFT of $x$
on a regular grid, and then interpolate the transform onto a polar
grid before 1D inverse Fourier transforms are applied to interpolated
points along the same radial lines.

The interpolation between the Cartesian and polar grid is the key
step in this procedure. It can be carried out using a ``gridding" algorithm 
that maintains the desired accuracy with low computational complexity. The 
gridding algorithm essentially allows us to perform a non-uniform FFT. 

The gridding operation requires the convolution between irregular
samples and a kernel calculated at regular sample position and vice
versa (Fig. \ref{fig:gridding} and \cite{jackson}).

A fast GPU implementation requires dividing the problem into each blocks with
similar computational loads.
Our approach starts with one bin containing all the points, and then 
recursively divides the most computationally intensive
bin into four equally sized bins. The computational intensity is
estimated by multiplying the number of samples in the bin by the number
of grid points in the bin. For the calculation in the GPU we assign
one thread block per bin.
The calculation strategy depends on the
number of samples in the bin. When the number of samples does not fit
in shared memory we assign all threads to each grid point and access
the samples in a coalesced way. When it does fit in shared memory we
assign one thread per grid point.

Further details of the code will be described in the
future. Preliminary tests indicate 50x speedup on a Tesla GPU C1060
compared to a fast CPU (e,g.: 3 msec for the FFTs, 6 msec for
gridding+inverse gridding for a 1024x1024 grid sampled with 180
angles, 1024 pixels per angle).

\subsection{Conclusions}
In summary we implemented a GPU-accelerated compressive phase contrast
tomography reconstruction using Basis Pursuit solvers for the high
throughput tomography beamline at the Advanced Light Source. The
reconstruction procedure was used to remove ring artifacts but it is
expected to also enable lower dose or smaller datasets for similar
image quality as shown for absorption-only (no phase contrast)
tomographic datasets exploting total variation regularization.

\subsection*{Acknowledgments} 
This work was supported by the Laboratory Directed Research and
Development Program of Lawrence Berkeley National Laboratory under
U.S. Department of Energy Contract No. DE-AC02-05CH11231.  We
acknowledge the use of the X-ray synchrotron micro-tomography beam
line (8.3.2) at the Advanced Light Source at LBNL, supported by the
Office of Science of the Department of Energy. Part of the work was
developed using Jacket GPU toolbox for matlab provided by Accelereyes.


\bibliographystyle{spiebib}   

\end{document}